\documentclass[11pt,draftclsnofoot,onecolumn]{IEEEtran}
\usepackage{cite}
\usepackage{graphicx}
\usepackage{psfrag}
\usepackage{subfigure}
\usepackage{url}
\usepackage{stfloats}
\usepackage{amsmath}
\usepackage{array}
\usepackage{amsmath}
\usepackage{amsfonts}
\usepackage{amssymb}
\usepackage{epsfig}
\usepackage{algorithm}
\usepackage{mathrsfs}
\usepackage{algpseudocode}
\usepackage{dsfont}

\newtheorem{theorem}{{Theorem}}
\newtheorem{definition}{{Definition}}
\newtheorem{remark}{{Remark}}

\begin{document}
\title{Linearly Coupled Communication Games}
\author{Yi~Su and~Mihaela van der Schaar% <-this % stops a space
%\thanks{M. Shell is with the Department
%of Electrical and Computer Engineering, Georgia Institute of Technology, Atlanta,
%\thanks{Manuscript received April 19, 2005; revised January 11, 2007.}
\\ Department of Electrical Engineering, UCLA}

%\markboth{Journal of \LaTeX\ Class Files,~Vol.~6, No.~1, January~2007}%
%{Shell \MakeLowercase{\textit{et al.}}: Bare Demo of IEEEtran.cls for Journals}


\maketitle

\begin{abstract}
%\boldmath
This paper discusses a special type of multi-user communication
scenario, in which users' utilities are linearly impacted by their
competitors' actions. First, we explicitly characterize the Nash
equilibrium and Pareto boundary of the achievable utility region.
Second, the price of anarchy incurred by the non-collaborative Nash
strategy is quantified. Third, to improve the performance in the
non-cooperative scenarios, we investigate the properties of an
alternative solution concept named conjectural equilibrium, in which
individual users compensate for their lack of information by forming
internal beliefs about their competitors. The global convergence of
the best response and Jacobi update dynamics that achieve various
conjectural equilibria are analyzed. It is shown that the Pareto
boundaries of the investigated linearly coupled games can be
sustained as stable conjectural equilibria if the belief functions
are properly initialized. The investigated models apply to a variety
of realistic applications encountered in the multiple access design,
including wireless random access and flow control.
\end{abstract}

\IEEEpeerreviewmaketitle

\section{Introduction}
Game theory provides a formal framework for studying the
interactions of strategic agents. Recently, there has been a surge
in research activities that employ game theory to model and analyze
a wide range of application scenarios in modern communication
networks \cite{Survery_Altman}-\cite{Survery_EPFL}. In communication
networks, any action taken by a single user usually affects the
utilities of the other users sharing the same resources. Depending
on the characteristics of different applications, numerous
game-theoretical models and solution concepts have been proposed to
describe the multi-user interactions and optimize the users'
decisions in communication networks. Roughly speaking, the existing
multi-user research can be categorized into two types,
non-cooperative games and cooperative games. Various game theoretic
solutions were developed to characterize the resulting performance
of the multi-user interaction, including the Nash Equilibrium (NE)
and the Pareto-optimality \cite{Game_book}.

Non-cooperative approaches generally assume that the participating
users simply choose actions to selfishly maximize their individual
utility functions. It is well-known that if devices operate in a
non-cooperative manner, this will generally limit their performance
as well as that of the whole system, because the available resources
are not always efficiently exploited due to the conflicts of
interest occurring among users \cite{Tragedy}. Most non-cooperative
approaches are devoted to investigating the existence and properties
of the NE. In particular, several non-cooperative game models, such
as S-modular games, congestion games, and potential games, have been
extensively applied in various communication scenarios
\cite{Sgame_1}-\cite{Pgame_1}. The price of anarchy, a measure of
how good the system performance is when users play selfishly and
reach the NE instead of playing to achieve the social optimum, has
also been addressed in several communication network applications
\cite{Johari}\cite{Roughgarden}.

On the other hand, cooperative approaches in communication theory
usually focus on studying how users can jointly improve their
performance when they cooperate. For example, the users may optimize
a common objective function, which represents the Pareto-optimal
social welfare allocation rule based on which the system-wide
resource allocation is performed \cite{Num}\cite{Zhu}. A profile of
actions is Pareto-optimal if there is no other profile of actions
that makes every player at least as well off and at least one player
strictly better off. Allocation rules, e.g. network utility
maximization, can provide reasonable allocation outcomes by
considering the trade-off between fairness and efficiency. Most
cooperative approaches focus on studying how to efficiently find the
optimum joint policy. It is worth mentioning that information
exchanges among users is generally required to enable users to
coordinate in order to achieve and sustain Pareto-efficient
outcomes.

In this paper, we present a game model for a particular type of
non-cooperative multi-user communication scenario. We name it
linearly coupled communication games, because users' utilities are
linearly impacted by their competitors' actions. In particular, the
main contributions of this paper are as follows. First, based on the
assumptions that we make about the properties of users' utility, we
characterize the inherent structures of the utility functions for
the linearly coupled games. Furthermore, based on the derived
utility forms, we explicitly quantify the NE and Pareto boundary for
the linearly coupled communication games. The price of anarchy
incurred by the selfish users playing the Nash strategy is
quantified. In addition, to improve the performance in the
non-cooperative scenarios, we investigate an alternative solution:
conjectural equilibrium (CE). Using this approach, individual users
are modeled as belief-forming agents that develop internal beliefs
about their competitors and behave optimally with respect to their
individual beliefs. Necessary and sufficient conditions that
guarantee the convergence of different dynamic update mechanisms,
including the best response and Jacobi update, are addressed. We
prove that these adjustment processes based on conjectures and
non-cooperative individual optimization can be globally driven to
Pareto-optimality in the linearly coupled games without the need of
real-time coordination information exchange among agents.

The rest of this paper is organized as follows. Section II defines
the linearly coupled communication games. For the investigated game
models, Section III explicitly computes the NE and Pareto boundary
of the achievable utility region and quantifies the price of
anarchy. Section IV introduces the CE and investigates its
properties under both the best response and Jacobi update dynamics.
Conclusions are drawn in Section V.

\section{Game Model}
In this section, we first provide a general game-theoretic
formulation of the multi-user interaction in communication systems.
Following the proposed definition, we define the linearly coupled
communication games and provide concrete examples of the
investigated game model.

\subsection{Linearly Coupled Communication Games}
The multi-user game in various communication scenarios can be
formally defined as a tuple
$\Gamma=\langle\mathcal{N},\mathcal{A},u,\mathcal{S},s\rangle$. In
particular, $\mathcal{N}=\{1,2,\ldots,N\}$ is the set of
communication devices, which are the rational decision-makers in the
system. Define $\mathcal{A}$ to be the joint action space
$\mathcal{A}=\times_{n\in\mathcal{N}} \mathcal{A}_n$, with
$\mathcal{A}_n$ being the action set available for user $n$. As
opposed to the traditional strategic game
definition\cite{Game_book}, we introduce two new elements
$\mathcal{S}$ and $s$ into the game formulation. Specifically,
$\mathcal{S}$ is the \emph{state space}
$\mathcal{S}=\times_{n\in\mathcal{N}} \mathcal{S}_n$, where
$\mathcal{S}_n\subseteq\mathcal{R}_+$ is the part of the state
relevant to user $n$. The state is defined to capture the effects of
the multi-user coupling such that each user's utility solely depends
on its own state and action. In other words, the utility function
$u=\times_{n\in\mathcal{N}}u_n$ is a mapping from the individual
users' state space and action space to real numbers, $u_n :
\mathcal{S}_n \times \mathcal{A}_n \rightarrow \mathcal{R}$. The
state determination function $s=\times_{n\in\mathcal{N}}s_n$ maps
joint actions to states for each component $s_n : \mathcal{A}
\rightarrow \mathcal{S}_n$. To capture the performance tradeoff, the
utility region is defined as
$\mathcal{U}=\{(u_1(\textbf{a}),\ldots,u_N(\textbf{a}))| \ \exists \
\textbf{a}=(a_1,a_2,\ldots,a_N)\in \mathcal{A}\}$.

\begin{definition}
A multi-user interaction is considered a \emph{linearly coupled
communication game} if the action set
$\mathcal{A}_n\subseteq\mathcal{R}_+$ is convex and the utility
function $u_n$ satisfies:
\begin{equation}
\label{eq:eqn1} u_n(\mathbf{a})=a_n^{\beta_n}\cdot s_n(\mathbf{a}),
\end{equation}
in which $\beta_n>0$. In particular, the basic assumptions about
$s_n(\mathbf{a})$ include:

\textbf{A1:} $s_n(\mathbf{a})$ is non-negative;

\textbf{A2:} Denote $s'_{nm}(\mathbf{a})=\frac{\partial
s_n(\mathbf{a})}{\partial a_m}$ and
$s''_{nm}(\mathbf{a})=\frac{\partial^2 s_n(\mathbf{a})}{\partial
a_m^2}$. $s_n(\mathbf{a})$ is strictly linear decreasing in $a_m,
\forall m\neq n$, i.e. $s'_{nm}(\mathbf{a})<0$ and
$s''_{nm}(\mathbf{a})=0$; $s_n(\mathbf{a})$ is non-increasing and
linear in $a_n$, i.e. $s'_{nn}(\mathbf{a})\leq0$ and
$s''_{nn}(\mathbf{a})=0$.

\textbf{A3:} $\frac{s_n(\mathbf{a})}{s'_{nm}(\mathbf{a})}$ is an
affine function, $\forall n \in \mathcal{N}\setminus \{m\}$.

\textbf{A4:}
$\frac{s'_{nm}(\mathbf{a})}{s_n(\mathbf{a})}=\frac{s'_{km}(\mathbf{a})}{s_k(\mathbf{a})},
\forall n,k\in \mathcal{N}\setminus \{m\}$;
$\frac{s'_{mm}(\mathbf{a})}{s_m(\mathbf{a})}=0$ or
$\frac{s'_{nm}(\mathbf{a})}{s_n(\mathbf{a})}, \ \forall n\neq m$.
\end{definition}

Assumptions A1 and A2 indicate that increasing $a_m$ for any $m\neq
n$ within the domain of $s_n(\mathbf{a})$ will linearly decrease
user $n$'s utility. Assumptions A3 and A4 imply that a user's action
has proportionally the same impact over the other users' utility.
The structure of the utility functions that satisfy assumptions
A1-A4 will be addressed in Section III.

\subsection{Illustrative Examples}
There are a number of multi-user communication scenarios that can be
modeled as linearly coupled communication games. For example, in the
random access scenario \cite{Yi_JSAC}, the action of a node is to
select its transmission probability and a node $n$ will
independently attempt transmission of a packet with transmit
probability $p_n$. The action set available to node $n$ is
$\mathcal{A}_n=[0,1]$ for all $n\in\mathcal{N}$. In this case, the
utility function is defined as
\begin{equation}
\label{eq:eqn2} u_n(\mathbf{p})=p_n\cdot \prod_{m\neq n}(1-p_m).
\end{equation}
As an additional example, in flow control \cite{Rate_Control}, $N$
Poisson streams of packets are serviced by a single exponential
server with departure rate $\mu$ and each class can adjust its
throughput $r_n$. The utility function is defined as the weighted
ratio of the throughput over the average experienced delay:
\begin{equation}
\label{eq:eqn3} u_n(\mathbf{r})=r_n^{\beta_n}\cdot (\mu-\sum_{m=1}^N
r_m),
\end{equation}
in which $\beta_n>0$ is interpreted as the weighting factor.
Specifically, we can see that the state determination functions are
$s_n(\mathbf{p})=\prod_{m\in \mathcal{N}\setminus \{n\}}(1-p_m)$ in
(\ref{eq:eqn2}) and $s_n(\mathbf{r})=\mu-\sum_{m=1}^N r_m$ in
(\ref{eq:eqn3}). It is straightforward to verify that these
functions satisfy assumptions A1-A4 for both (\ref{eq:eqn2}) and
(\ref{eq:eqn3}).

In this paper, we are interested in comparing the achievable
performance attained by different game-theoretic solution concepts.
On one hand, it is well-known that NE is generally inefficient in
communication games \cite{NE_inef}, but it may not require explicit
message exchanges, while Pareto-optimality can usually be achieved
only by exchanging implicit or explicit coordination messages among
the participating users. On the other hand, in several recent works
\cite{Yi_TSP}\cite{Yi_JSAC}, we have applied an alternative solution
in different communication scenarios to improve the system
performance in non-cooperative settings, namely the conjectural
equilibrium \cite{CE}. The following sections aim to compare the
solutions of NE, Pareto boundary, and CE in terms of the payoffs and
informational requirements in the linearly coupled multi-user
interaction satisfying the assumptions A1-A4.

%In our previous work, we have applied the conjectural equilibrium in
%the context of random access scenario and show that all the
%operating points in the throughput region are essentially stable
%conjectural equilibrium. This paper aims to extend the previous
%results about conjectural equilibrium to a general type of
%multi-user interactions where users are mutually linearly coupled.

\section{Computation of the Nash Equilibrium and Pareto Boundary for Linearly Coupled Games}
In this section, we show that the computation of the NE and the
Pareto boundary in linearly coupled games is equivalent to solving
linear equations. Specifically, we investigate the inherent
structures of the utility functions satisfying assumptions A1-A4 and
define two basic types of linearly coupled games. The performance
loss incurred by the Nash strategy are quantified for Type II games.

\subsection{Nash Equilibrium}
In non-cooperative games, the participating users simply choose
actions to selfishly maximize their individual utility functions.
The steady state outcome of such interactions is an operating point,
at which given the other users' actions, no user can increase its
utility alone by unilaterally changing its action. This operating
point is known as the Nash equilibrium, which is formally defined
below \cite{Game_book}.
\begin{definition}
A profile $ \mathbf{a}$ of actions constitutes a \emph{Nash
equilibrium} of $\Gamma$ if $u_n(a_n,\mathbf{a}_{-n})\geq
u_n(a'_n,\mathbf{a}_{-n})$ for all $a'_n\in \mathcal{A}_n$ and
$n\in\mathcal{N}$.
\end{definition}

We are interested in computing the NE in the linear coupled games.
From equation (\ref{eq:eqn1}), we have
\begin{equation}
\label{eq:eqn4} \frac{\partial \log[u_n(\mathbf{a})]}{\partial
a_m}=\left\{
       \begin{array}{ll}
         \beta_n/a_n+s'_{nn}(\mathbf{a})/s_n(\mathbf{a}), & if \ m=n;\\
         s'_{nm}(\mathbf{a})/s_n(\mathbf{a}), & otherwise.
       \end{array}
     \right.
\end{equation}

On one hand, if $s'_{nn}(\mathbf{a})=0, \forall n \in \mathcal{N}$,
since user $n$'s utility function strictly increases in $a_n$, we
have trivial NE at which $a_n^*$ is the maximal element in
$\mathcal{A}_n$ that lies in the domain of $s(\cdot)$, $\forall n
\in \mathcal{N}$.

On the other hand, if $s'_{nn}(\mathbf{a})\neq0, \forall n \in
\mathcal{N}$, according to assumption A3, since the multi-user
interactions are linearly coupled, we have
\begin{equation}
\label{eq:eqn5}
s_n(\mathbf{a})=f_n^m(\mathbf{a}_{-m})+g_n^m(\mathbf{a}_{-m})a_m,
\end{equation}
where $f_n^m(\mathbf{a}_{-m}), g_n^m(\mathbf{a}_{-m})$ are both
polynomials and $g_n^n(\mathbf{a}_{-n})\neq0$. From this, it follows
\begin{equation}
\label{eq:eqn6}
\frac{s'_{nn}(\mathbf{a})}{s_n(\mathbf{a})}=\bigg[\frac{f_n^n(\mathbf{a}_{-n})}{g_n^n(\mathbf{a}_{-n})}+a_n\bigg]^{-1}.
\end{equation}
At NE, we have
\begin{equation}
\label{eq:eqn7} \frac{\partial \log[u_n(\mathbf{a})]}{\partial
a_n}=0, \forall n \in \mathcal{N}.
\end{equation}
Under assumption A3 and A4,
$\frac{f_n^n(\mathbf{a}_{-n})}{g_n^n(\mathbf{a}_{-n})}$ is a affine
function, which enables us to explicitly characterize the NE. Denote
$\frac{f_n^n(\mathbf{a}_{-n})}{g_n^n(\mathbf{a}_{-n})}=h_n(\mathbf{a}_{-n})$.
Equation (\ref{eq:eqn7}) can be rewritten as
\begin{equation}
\label{eq:eqns}\beta_n\cdot h_n(\mathbf{a}_{-n})+(\beta_n+1)\cdot
a_n=0, \forall n \in \mathcal{N}.
\end{equation}
Therefore, the solutions of Equations (\ref{eq:eqns}) are the NE of
the linearly coupled games and computing the NE is equivalent to
solving $N$-dimension linear equations. The following theorem
indicates the inherent structure of the utility functions
$\{u_n\}_{n=1}^N$ when the requirements A1-A3 are satisfied.
\begin{theorem} \label{th:th1} Under assumptions A1-A3, the irreducible factors of $s_n(\mathbf{a})$ over the integers are affine functions and have no variables in common.
\end{theorem}

\emph{Proof}: Denote the factorization of $s_n(\mathbf{a})$ as
\begin{equation}
\label{eq:eqn9} s_n(\mathbf{a})=\prod_{i=1}^{M_n}b_n^i(\mathbf{a}),
\end{equation}
in which $M_n$ represents the number of the non-constant irreducible
factors in $s_n(\mathbf{a})$. Define $\mathrm{V}(\cdot)$ as the
mapping from a polynomial to the set of variables that appear in
that polynomial. Based on assumption A2, we immediately have
\begin{displaymath}
\mathrm{V}(b_n^i(\mathbf{a})) \cap \mathrm{V}(b_n^j(\mathbf{a})) =
\varnothing, \forall i,j(j\neq i),n.
\end{displaymath}

Without loss of generality, we assume that $a_j \in
\mathrm{V}(b_n^1(\mathbf{a}))$ and
$b_n^1(\mathbf{a})=f_{b_n^1}^j(\mathbf{a}_{-j})+g_{b_n^1}^j(\mathbf{a}_{-j})a_j$.
Then $f_n^j(\mathbf{a}_{-j}), g_n^j(\mathbf{a}_{-j})$ in
(\ref{eq:eqn5}) are given by
\begin{displaymath}
f_n^j(\mathbf{a}_{-j})=f_{b_n^1}^j(\mathbf{a}_{-j})\cdot
\prod_{i=2}^{M_n}b_n^i(\mathbf{a}), \ \textrm{and} \
g_n^j(\mathbf{a}_{-j})=g_{b_n^1}^j(\mathbf{a}_{-j})\cdot
\prod_{i=2}^{M_n}b_n^i(\mathbf{a}).
\end{displaymath}
Therefore,
$\frac{f_n^m(\mathbf{a}_{-m})}{g_n^m(\mathbf{a}_{-m})}=\frac{f_{b_n^1}^j(\mathbf{a}_{-j})}{g_{b_n^1}^j(\mathbf{a}_{-j})}$.
By assumption A3, we have that the degree of
$\frac{f_{b_n^1}^j(\mathbf{a}_{-j})}{g_{b_n^1}^j(\mathbf{a}_{-j})}$
is less than or equal to 1. Since $b_n^1(\mathbf{a})$ is
irreducible, we can conclude that $g_{b_n^1}^j(\mathbf{a}_{-j})$ is
a constant and the degree of $f_{b_n^1}^j(\mathbf{a}_{-j})$ is less
than or equal to 1. Note that the arguments above hold, $\forall
j,n$. Therefore, the degree of $b_n^i(\mathbf{a})$ is one, $\forall
n\in \mathcal{N},i=1,\ldots,M_n$, which concludes the proof.
$\blacksquare$

\subsection{Pareto Boundary}
Since $\log(\cdot)$ is concave and $\log[u_n(\mathbf{a})]$ is a
composition of affine functions \cite{convex_book},
$u_n(\mathbf{a})$ is log-concave in $\mathbf{a}$ and the log-utility
region $\log \mathcal{U}$ is convex. Therefore, we can characterize
the Pareto boundary of the utility region as a set of $\textbf{a}$
optimizing the following weighted proportional fairness
objective\footnote[1]{Note that the utility region $\mathcal{U}$ is
not necessarily convex. Therefore, its Pareto boundary may not be
characterized by the weighted sum of
$\{u_n(\mathbf{a})\}_{n=1}^N$.}:
\begin{equation}
\label{eq:eqnop} \max_{\textbf{a}}
\sum_{n=1}^N\omega_n\log[u_n(\mathbf{a})],
\end{equation}
for all possible sets of $\{\omega_n\}$ satisfying $\omega_n\geq0$
and $\sum_{n=1}^N\omega_n=1$. Denote the optimal solution of problem
(\ref{eq:eqnop}) as $\textbf{a}^{PB}$, which satisfies the following
first-order condition:
\begin{equation}
\label{eq:eqn8} \frac{\partial
\sum_{k=1}^N\omega_k\log[u_k(\mathbf{a})]}{\partial
a_n}\biggr\rvert_{\textbf{a}=\textbf{a}^{PB}}=0, \forall n \in
\mathcal{N},
\end{equation}

Under assumptions A1-A3, the LHS of equation (\ref{eq:eqn8}) can be
rewritten as
\begin{equation} \label{eq:eqn10} \frac{\partial
\sum_{k=1}^N\omega_k\log[u_k(\mathbf{a})]}{\partial
a_m}=\omega_m\bigg(\frac{\beta_m}{a_m}+\frac{s'_{mm}(\mathbf{a})}{s_m(\mathbf{a})}\bigg)+\sum_{k\neq
m}\omega_k\frac{s'_{km}(\mathbf{a})}{s_k(\mathbf{a})}.
\end{equation}
By Theorem \ref{th:th1} and assumption A4, we have
\begin{equation} \label{eq:eqn11}
\frac{s'_{km}(\mathbf{a})}{s_k(\mathbf{a})}=\frac{1}{\psi_m(\mathbf{a})},
\ \forall k\in \mathcal{N}\setminus \{m\},
\end{equation}
in which $\psi_m(\mathbf{a})$ is a affine function. Therefore,
equation (\ref{eq:eqn10}) is equivalent to
\begin{equation}
\label{eq:eqn12} \frac{\partial
\sum_{k=1}^N\omega_k\log[u_k(\mathbf{a})]}{\partial a_m}=\left\{
       \begin{array}{ll}
         \beta_m\omega_m/a_m+(1-\omega_m)/\psi_m(\mathbf{a}), & if \ s'_{mm}(\mathbf{a})=0;\\
         \beta_m\omega_m/a_m+1/\psi_m(\mathbf{a}), & otherwise.
       \end{array}
     \right.
\end{equation}
We can compute the Pareto boundary of the linearly coupled games by
solving linear equations:
\begin{equation}
\label{eq:eqn13} \frac{\partial
\sum_{k=1}^N\omega_k\log[u_k(\mathbf{a})]}{\partial a_m}=0
\Rightarrow \left\{
       \begin{array}{ll}
         \beta_m\omega_m\psi_m(\mathbf{a})+(1-\omega_m)a_m=0, & if \ s'_{mm}(\mathbf{a})=0;\\
         \beta_m\omega_m\psi_m(\mathbf{a})+a_m=0, & otherwise.
       \end{array}
     \right.
\end{equation}

Theorem \ref{th:th1} reveals the structural properties of the
utility functions $\{u_n\}_{n=1}^N$ when assumption A1-A3 are
satisfied. Based on Theorem \ref{th:th1}, the following theorem
further refines these properties of $\{u_n\}_{n=1}^N$ when the
additional assumption A4 is imposed.
\begin{theorem} \label{th:th2} Under assumptions A1-A4, for any polynomial $b_n^i(\mathbf{a})$ in the
factorization $s_n(\mathbf{a})=\prod_{i=1}^{M_n}b_n^i(\mathbf{a})$,
$\forall n \in \mathcal{N}$, if
$|\mathrm{V}(b_n^i(\mathbf{a}))|\geq2$ or
$\mathrm{V}(b_n^i(\mathbf{a}))=\{a_n\}$, $b_n^i(\mathbf{a})$ is an
irreducible factor of $s_m(\mathbf{a})$, $\forall m \in
\mathcal{N}$; if $\mathrm{V}(b_n^i(\mathbf{a}))=\{a_m\}, m\neq n$,
$b_n^i(\mathbf{a})$ is an irreducible factor of $s_j(\mathbf{a})$,
$\forall j \in \mathcal{N}/\{m\}$.
\end{theorem}

\emph{Proof}: By assumption A2, $s'_{nm}(\mathbf{a})<0, \forall
m\neq n$, we have $|\mathrm{V}(s_n(\mathbf{a}))|\geq N-1, \forall n
\in \mathcal{N}$. By Theorem \ref{th:th1}, the irreducible factors
of $s_n(\mathbf{a})$ have no common variables and they are affine
functions. Suppose $|\mathrm{V}(b_n^i(\mathbf{a}))|\geq2$ and
$\{a_m,a_l\}\in \mathrm{V}(b_n^i(\mathbf{a})$. By assumption A4, we
know that
$\frac{s'_{nm}(\mathbf{a})}{s_n(\mathbf{a})}=\frac{s'_{km}(\mathbf{a})}{s_k(\mathbf{a})}=\frac{b_{nm}'^i(\mathbf{a})}{b_n^i(\mathbf{a})},
\forall n,k\in \mathcal{N}\setminus \{m\}$. Therefore, it follows
\begin{equation}
\label{eq:eqn14}
s_k(\mathbf{a})=\frac{s'_{km}(\mathbf{a})b_n^i(\mathbf{a})}{b_{nm}'^i(\mathbf{a})}.
\end{equation}
Since $b_{nm}'^i(\mathbf{a})$ is a constant, we can see that
$b_n^i(\mathbf{a})$ is an irreducible factor of $s_k(\mathbf{a})$,
$\forall k \in \mathcal{N}\setminus \{m\}$. By symmetry, we can
conclude that $b_n^i(\mathbf{a})$ must also be an irreducible factor
of $s_k(\mathbf{a})$, $\forall k \in \mathcal{N}\setminus \{l\}$.
Therefore, $b_n^i(\mathbf{a})$ is an irreducible factor of
$s_k(\mathbf{a})$, $\forall k \in \mathcal{N}$. Similarly, we can
prove the remaining parts of Theorem \ref{th:th2}. $\blacksquare$

\begin{remark} \label{rm:rm1}
For the linearly coupled games satisfying assumptions A1-A4, suppose
we factorize all users' state functions. Theorem \ref{th:th2}
indicates that any factor with at least two variables must be a
common factor of all the users' state functions, and any factor with
a single variable $a_k$ must be a common factor of state functions
for users excluding $k$. In reality, it corresponds to the
communication scenarios in which the state, i.e. the multi-user
coupling, is impacted by a set of users that result in a similar
signal to all the users.

%For example, the utility functions in a three-player game satisfy
%A1-A4: $u_1(\mathbf{a})=a_1^2\cdot(2-a_2-a_3),
%u_2(\mathbf{a})=a_2^{0.5}\cdot(1-a_1)(2-a_2-a_3),
%u_3(\mathbf{a})=a_3^{1.5}\cdot(1-a_1)(2-a_2-a_3).$ We can see that,
%$|\mathrm{V}(2-a_2-a_3)|=2$ and $2-a_2-a_3$ is a common factor of
%$s_n(\mathbf{a}), n=1,2,3$. On the other hand,
%$|\mathrm{V}(1-a_1)|=1$ and $1-a_1$ is a common factor of
%$s_2(\mathbf{a})$ and $s_3(\mathbf{a})$.

We define two basic types of linearly coupled games satisfying the
assumptions A1-A4. In Type I games, user $k$'s action linearly
decreases all the users' states but itself. Hence, the utility
functions take the form
\begin{equation}
\label{eq:eqn15} u_n(\mathbf{a})=a_n^{\beta_n}\cdot \prod_{m\neq
n}(\mu_m-\tau_ma_m).
\end{equation}
In Type II games, all the users share the same non-factorizable
state function and their utility functions are given by
\begin{equation}
\label{eq:eqn16} u_n(\mathbf{a})=a_n^{\beta_n}\cdot
(\mu-\sum_{m=1}^N\tau_ma_m).
\end{equation}
\end{remark}
As special examples, the random access problem in (\ref{eq:eqn2})
belongs to Type I games and the rate control problem in
(\ref{eq:eqn3}) belongs to Type II games. In fact, all the games
that have the properties A1-A4 can be viewed as compositions of
these two basic types of games (See the example in Remark
\ref{rm:rm1}). Therefore, investigating the two basic types provides
us the fundamental understanding of the linearly coupled multi-user
interaction. A brief summary of the properties of Type I games will
be provided in Section IV.E. For the details about its various
game-theoretic solutions, we refer the readers to \cite{Yi_JSAC} and
the references therein. The rest of this paper will focus on Type II
games.

\subsection{Nash Equilibrium and Pareto Boundary in Type II Games}
For Type II games with utility functions given in (\ref{eq:eqn16}),
we have
\begin{equation}
\label{eq:eqn17}
\frac{s'_{nn}(\mathbf{a})}{s_n(\mathbf{a})}=\frac{-\tau_n}{\mu-\sum_{m=1}^N\tau_ma_m}.
\end{equation}
Therefore, Equation (\ref{eq:eqns}) can be reduced to
\begin{equation}
\label{eq:eqn18}(1+\beta_n)\tau_na_n+\beta_n\sum_{m\neq
n}\tau_ma_m=\beta_n\mu, \forall n \in \mathcal{N}.
\end{equation}
The solution of the linear equations gives the NE, and its closed
form has been addressed in \cite{NE_RC} for $\tau_n=1, \forall n\in
\mathcal{N}$. For the general case, it is easy to verify that the NE
is given by
\begin{equation}
\label{eq:eqn19}
a_n^{NE}=\frac{\beta_n\mu}{\tau_n(1+\sum_{m=1}^N\beta_m)},  \forall
n \in \mathcal{N}.
\end{equation}

Similarly, to compute the Pareto boundary of Type II games, Equation
(\ref{eq:eqn12}) can be reduced to
\begin{equation}
\label{eq:eqn20}(1+\omega_n\beta_n)\tau_na_n+\omega_n\beta_n\sum_{m\neq
n}\tau_ma_m=\omega_n\beta_n\mu, \forall n \in \mathcal{N}.
\end{equation}
The solution is given by
\begin{equation}
\label{eq:eqn21}
a_n^{PB}=\frac{\omega_n\beta_n\mu}{\tau_n(1+\sum_{m=1}^N\omega_m\beta_m)},
\forall n \in \mathcal{N}.
\end{equation}

From Section II.B, we know that the region $\log\mathcal{U}$ is
convex. Therefore, we can compare the efficiency of
$\mathbf{a}^{NE}$ and $\mathbf{a}^{PB}$ using the system-utility
metric $\sum_{n=1}^N\omega_n\log[u_n(\mathbf{a})]$. Specifically, we
have
\begin{equation}
\label{eq:eqn22}
\sum_{n=1}^N\omega_n\log\frac{u_n(\mathbf{a}^{NE})}{u_n(\mathbf{a}^{PB})}=\sum_{n=1}^N\omega_n\beta_n\log\frac{1+\sum_{j=1}^N\omega_j\beta_j}{\omega_n(1+\sum_{j=1}^N\beta_j)}+\log\frac{1+\sum_{j=1}^N\omega_j\beta_j}{1+\sum_{j=1}^N\beta_j}.
\end{equation}
Denote $w_0=1$,
$x_0=\frac{1+\sum_{j=1}^N\omega_j\beta_j}{1+\sum_{j=1}^N\beta_j}$,
$w_n=\omega_n\beta_n$, and
$x_n=\frac{1+\sum_{j=1}^N\omega_j\beta_j}{\omega_n(1+\sum_{j=1}^N\beta_j)},
\forall n \in \mathcal{N}$. Therefore,
\begin{equation}
\label{eq:eqn23}
\sum_{n=1}^N\omega_n\log\frac{u_n(\mathbf{a}^{NE})}{u_n(\mathbf{a}^{PB})}=\sum_{n=1}^Nw_n\log{x_n}+w_0\log{x_0}=\sum_{n=0}^Nw_n\cdot\log{(\prod_{n=0}^Nx_n^{w_n})^{1/\sum_{n=0}^Nw_n}}.
\end{equation}
Using the inequalities among the arithmetic, geometric and harmonic
means \cite{Handbook}, we have
\begin{equation}
\label{eq:eqn24}
\frac{(1+\sum_{n=1}^N\omega_n\beta_n)^2}{(1+\sum_{n=1}^N\omega_n^2\beta_n)(1+\sum_{n=1}^N\beta_n)}=\frac{\sum_{n=0}^Nw_n}{\sum_{n=0}^N\frac{w_n}{x_n}}\leq\bigl(\prod_{n=0}^Nx_n^{w_n}\bigr)^{\frac{1}{\sum_{n=0}^Nw_n}}\leq\frac{\sum_{n=0}^Nx_nw_n}{\sum_{n=0}^Nw_n}=1.
\end{equation}
Both inequalities hold with equality if and only if
$x_0=x_1=\ldots=x_N$, i.e. $\omega_1=\ldots=\omega_N=1$. However,
since we require $\sum_{n=1}^N\omega_n=1$, (\ref{eq:eqn24}) holds as
strict inequalities, which leads to
\begin{equation}
\label{eq:eqn25}
(1+\sum_{n=1}^N\omega_n\beta_n)\cdot\log{\frac{(1+\sum_{n=1}^N\omega_n\beta_n)^2}{(1+\sum_{n=1}^N\omega_n^2\beta_n)(1+\sum_{n=1}^N\beta_n)}}<\sum_{n=1}^N\omega_n\log\frac{u_n(\mathbf{a}^{NE})}{u_n(\mathbf{a}^{PB})}<0.
\end{equation}
Based on Equation (\ref{eq:eqn25}), we can make two important
observations. First, due to the lack of coordination, the NE in Type
II games is always strictly Pareto inefficient. Second, as opposed
to Type I games where NE may result in zero utility for certain
users \cite{Yi_JSAC}, the efficiency loss in Type II games are lower
bounded, which means that every user receives positive payoff at NE.
Noticing that the performance gap between $u_n(\mathbf{a}^{NE})$ and
$u_n(\mathbf{a}^{PB})$ is non-zero, we will investigate how the
non-cooperative CE solution can improve the system performance for
Type II games.

\section{Conjectural Equilibrium for the Linearly Coupled Games}
\subsection{Definitions}
In game-theoretic analysis, conclusions about the reached equilibria
are based on assumptions about what knowledge the players possess.
For example, the standard NE strategy assumes that every player
believes that the other players' actions will not change at NE.
Therefore, it chooses to myopically maximize its immediate payoff
\cite{Game_book}. Therefore, the players operating at equilibrium
can be viewed as decision makers behaving optimally with respect to
their \emph{beliefs} about the strategies of other players.

To avoid detrimental Nash strategy and encourage cooperation, the
conjecture-based model has been introduced by Wellman and others
\cite{Wellman}\cite{CE} to enable non-cooperative players to build
belief models about how their competitors' reactions vary in
response to their own action changes. Specifically, each player has
some belief about the state that would result from performing its
available actions. The \emph{belief function} $\tilde{s}_n$ is
defined to be $\tilde{s}_n:\mathcal{A}_n\rightarrow\mathcal{S}_n$
such that $\tilde{s}_n(a_n)$ represents the state that player $n$
believes it would result in if it selects action $a_n$ . Notice that
the beliefs are not expressed in terms of other players' actions and
preferences, and the multi-user coupling in these beliefs is
captured indirectly by individual players forming conjectures of the
effects of their own actions. By deploying such a behavior model,
players will no longer adopt myopic behaviors that do not forecast
$\tilde{s}_n$, but rather they will form beliefs $\tilde{s}_n(a_n)$
about how their actions $a_n$ will influence the aggregate effects
$\tilde{s}_n$ incurred by their competitors' responses and, based on
these beliefs, they will choose the action $a_n\in \mathcal{A}_n$ if
it believes that this action will maximize its utility. The steady
state of such a play among belief-forming agents can be
characterized as a conjectural equilibria.

\begin{definition}
In the game $\Gamma$, a configuration of belief functions
$(\tilde{s}_1^*,\ldots,\tilde{s}_N^*)$ and a joint action
$a^*=(a_1^*,\ldots,a_N^*)$ constitute a conjectural equilibrium, if
for each $n\in \mathcal{N}$,
\begin{displaymath} \tilde{s}_n^*(a_n^*)=s_n(a_1^*,\ldots,a_N^*) \ \textrm{and} \ a_n^*=\arg \max_{a_n\in \mathcal{A}_n}u_n(\tilde{s}_n^*(a_n),a_n).
\end{displaymath}
\end{definition}

From the above definition, we can see that, at CE, all players'
expectations based on their beliefs are realized and each agent
behaves optimally according to its expectation. In other words,
agents' beliefs are consistent with the outcome of the play and they
use ``conjectured best responses" in their individual optimization
program. The key challenges are how to configure the belief
functions such that cooperation can be sustained in such a
non-cooperative setting and how to design the evolution rules such
that the communication system can dynamically converge to a CE
having satisfactory performance.

\subsection{Linear Beliefs}
As discussed before, the belief functions need to be defined in
order to investigate the existence of CE. To define the belief
functions, we need to express agent $n$'s expected state
$\tilde{s}_n$ as a function of its own action $a_n$. The simplest
approach is to design linear belief models for each user, i.e.
player $n$'s belief function takes the form
\begin{equation}
\label{eq:eqn26}
\tilde{s}_n(a_n)=\bar{s}_n-\lambda_n(a_n-\bar{a}_n),
\end{equation}
for $n\in \mathcal{N}$. The values of $\bar{s}_n$ and $\bar{a}_n$
are specific states and actions, called \emph{reference points} and
$\lambda_n$ is a positive scalar. In other words, user $n$ assumes
that other players will observe its deviation from its reference
point $\bar{a}_n$ and the aggregate state deviates from the
reference point $\bar{s}_n$ by a quantity proportional to the
deviation of $a_n-\bar{a}_n$. How to configure $\bar{s}_n,
\bar{a}_n$, and $\lambda_n$ will be addressed in the rest of this
paper. We focus on the linear belief represented in
(\ref{eq:eqn26}), because this simple belief form is sufficient to
drive the resulting non-cooperative equilibrium to the Pareto
boundary.

The goal of user $n$ is to maximize its expected utility
 $a_n^{\beta_n}\cdot\tilde{s}_n(a_n)$ taking into account the conjectures that it
has made about the other users. Therefore, the optimization a user
needs to solve becomes:
\begin{equation}
\label{eq:eqn27} \max_{a_n\in
\mathcal{A}_n}a_n^{\beta_n}\cdot\Big[\bar{s}_n-\lambda_n(a_n-\bar{a}_n)\Big].
\end{equation}
For $\lambda_k>0$, user $n$ believes that increasing $a_n$ will
further reduce its conjectured state $\bar{s}_n$. The optimal
solution of (\ref{eq:eqn27}) is given by
\begin{equation}
\label{eq:eqn28} a_n^*= \frac
{\beta_n(\bar{s}_n+\lambda_n\bar{a}_n)}{\lambda_n(1+\beta_n)}.
\end{equation}

In the following, we first show that forming simple linear beliefs
in (\ref{eq:eqn26}) can cause all the operating points in the
achievable utility region to be CE.
\begin{theorem}
\label{th:th3} For Type II games, all the positive operating points
in the utility region $\mathcal{U}$ are essentially CE.
\end{theorem}

\emph{Proof}: For each positive operating point
$(u_1^*,\ldots,u_N^*)$ (i.e. $u_n^*>0, \forall n \in \mathcal{N}$)
in the utility region $\mathcal{U}$, there exists at least one joint
action profile $(a^*_1,\ldots,a^*_N)\in \mathcal{A}$ such that
$u_n^*=u_n(\textbf{a}^*)$, $\forall n\in\mathcal{N}$. We consider
setting the parameters in the belief functions
$\{\tilde{s}_n(a_n)\}_{n=1}^N$ to be:
\begin{equation}
\label{eq:eqn29}
\lambda_n^*=\beta_n\cdot\frac{\mu-\sum_{m=1}^N\tau_ma_m^*}{a_n^*},
\forall n \in \mathcal{N}.
\end{equation}
It is easy to check that, if the reference points are
$\bar{s}_n=\mu-\sum_{m=1}^N\tau_ma_m^*,\bar{a}_n=a_n^*$, we have
$\tilde{s}_n(a_n^*)=s_n(a_1^*,\ldots,a_N^*)$ and $a_n^*=\arg
\max_{a_n\in \mathcal{A}_n}u_n(\tilde{s}_n(a_n),a_n)$. Therefore,
this belief function configuration and the joint action
$\textbf{a}^*=(a_1^*,\ldots,a_N^*)$ constitute the CE that results
in the utility $(u_1^*,\ldots,u_N^*)$.\ $\blacksquare$

Theorem \ref{th:th3} establishes the existence of CE, i.e. for a
particular $\textbf{a}^*\in \mathcal{A}$, how to choose the
parameters $\{\bar{s}_n,\bar{a}_n,\lambda_n\}_{n=1}^N$ such that
$\textbf{a}^*$ is a CE. However, it neither tells us how these CE
can be achieved and sustained in the dynamic setting nor clarifies
how different belief configurations can lead to various CE.

We consider the dynamic scenarios in which users revise their
reference points based on their past local observations over time.
Let $s_n^t,a_n^t,\tilde{s}_n^t,\bar{s}_n^t,\bar{a}_n^t$ be user
$n$'s state, action, belief function, and reference points at stage
$t$, in which $s_n^t=\mu-\sum_{m=1}^N\tau_ma_m^t$. We propose a
simple rule for individual users to update their reference points.
At stage $t$, user $n$ sets its $\bar{s}_n^t$ and $\bar{a}_n^t$ to
be $s_n^{t-1}$ and $a_n^{t-1}$. In other words, user $n$'s
conjectured utility function at stage $t$ is
\begin{equation}
\label{eq:eqn30}
u_n^t(\tilde{s}_n^t(a_n),a_n)=a_n^{\beta_n}\cdot\Bigl[\mu-\sum_{m=1}^N\tau_ma_m^{t-1}-\lambda_n(a_n-a_n^{t-1})\Bigr].
\end{equation}
Since we have defined the users' utility function at stage $t$, upon
specifying the rule of how user $n$ updates its action $a_n^t$ based
on its utility function $u_n^t(\tilde{s}_n^t(a_n),a_n)$, the
trajectory of the entire dynamic process is determined. The
remainder of this paper will investigate the dynamic properties of
the best response and Jacobi update mechanisms and the performance
trade-off among the competing users at the resulting steady-state
CE. In particular, for fixed $\{\lambda_n\}_{n=1}^N$, Section IV-C
derives necessary and sufficient conditions for the convergence of
the best response and the Jacobi update dynamics. Section IV-D
quantitatively describes the limiting CE for given
$\{\lambda_n\}_{n=1}^N$ and investigates how the parameters
$\{\lambda_n\}_{n=1}^N$ should be properly chosen such that Pareto
efficiency can be achieved.

\subsection{Dynamic Algorithms}
\subsubsection{Best Response}In the best response algorithm, each
user updates its action using the best response that maximizes its
conjectured utility function in (\ref{eq:eqn30}). Therefore, at
stage $t$, user $n$ chooses its action according to
\begin{equation}
\label{eq:eqn31} a_n^t=B_n(\textbf{a}^{t-1}):= \frac
{\beta_n(\mu-\sum_{m \in
\mathcal{N}\setminus\{n\}}\tau_ma_m^{t-1})}{\lambda_n(1+\beta_n)}+\frac
{\beta_n(\lambda_n-\tau_n)a_n^{t-1}}{\lambda_n(1+\beta_n)}.
\end{equation}
We are interested in characterizing the convergence of the update
mechanism defined by (\ref{eq:eqn31}) when using various $\lambda_n$
to initialize the belief function $\tilde{s}_n$.

To analyze the convergence of the best response dynamics, we
consider the Jacobian matrix of the self-mapping function in
(\ref{eq:eqn31}). Let $J_{ik}$ denote the element at row $i$ and
column $k$ of the Jacobian matrix $\textbf{J}$. The elements of the
Jacobian matrix $\textbf{J}^{BR}$ of (\ref{eq:eqn31}) are defined
as:
\begin{equation}
\label{eq:eqn32} J^{BR}_{ik}=\frac{\partial a_i^t}{\partial
a_k^{t-1}}=\left\{
\begin{array}{cl}
\frac
{\beta_k(\lambda_k-\tau_k)}{\lambda_k(1+\beta_k)},& \text{if $i=k$},\\
-\frac {\beta_i\tau_k}{\lambda_i(1+\beta_i)},& \text{if $i\neq
k$}.\end{array} \right.
\end{equation}
For Type II games, the following theorem gives a necessary and
sufficient condition under which the best response dynamics defined
in (\ref{eq:eqn31}) converges.

\begin{theorem}
\label{th:th4} For Type II games, a necessary and sufficient
condition for the best response dynamics to converge is
\begin{equation}
\label{eq:eqn33}
\sum_{n=1}^N\frac{\tau_n\beta_n}{\lambda_n(1+2\beta_n)}<1.
\end{equation}
\end{theorem}

\emph{Proof}: The best response dynamics converges if and only if
the eigenvalues $\{\xi_n^{BR}\}_{n=1}^{N}$ of the Jacobian matrix
$\textbf{J}^{BR}$ in (\ref{eq:eqn32}) are all inside the unit circle
of the complex plane \cite{Contraction_mapping}, i.e.
$|\xi_n^{BR}|<1, \forall n \in \mathcal{N}$. To determine the
eigenvalues of $\textbf{J}^{BR}$, we have
\begin{displaymath}
\begin{split}\det(\xi I-\textbf{J}^{BR})&=\left|
                              \begin{array}{cccc}
                                \xi-\frac
{\beta_1(\lambda_1-\tau_1)}{\lambda_1(1+\beta_1)} & \frac {\beta_1\tau_2}{\lambda_1(1+\beta_1)} & \ldots & \frac {\beta_1\tau_N}{\lambda_1(1+\beta_1)} \\
                                \frac {\beta_2\tau_1}{\lambda_2(1+\beta_2)} & \xi-\frac
{\beta_2(\lambda_2-\tau_2)}{\lambda_2(1+\beta_2)} & \ldots & \frac {\beta_2\tau_N}{\lambda_2(1+\beta_2)} \\
                                \vdots & \vdots & \ddots & \vdots \\
                                \frac {\beta_N\tau_1}{\lambda_N(1+\beta_N)} & \frac {\beta_N\tau_2}{\lambda_N(1+\beta_N)} & \ldots & \xi-\frac
{\beta_N(\lambda_N-\tau_N)}{\lambda_N(1+\beta_N)} \\
                              \end{array}
                            \right| \\
&=\left|
                              \begin{array}{cccc}
                                \xi-\frac
{\beta_1(\lambda_1-\tau_1)}{\lambda_1(1+\beta_1)} & \frac {\tau_2}{\tau_1}\bigl(\frac{\beta_1}{1+\beta_1}-\xi\bigr) & \ldots & \frac {\tau_N}{\tau_1}\bigl(\frac{\beta_1}{1+\beta_1}-\xi\bigr) \\
                                \frac {\beta_2\tau_1}{\lambda_2(1+\beta_2)} & \xi-\frac
{\beta_2}{1+\beta_2} & \ldots & 0 \\
                                \vdots & \vdots & \ddots & \vdots \\
                                \frac {\beta_N\tau_1}{\lambda_N(1+\beta_N)} & 0 & \ldots & \xi-\frac
{\beta_N}{1+\beta_N} \\
                              \end{array}
                            \right| \\
&=\left|
                              \begin{array}{cccc}
                                \bigl(\xi-\frac{\beta_1}{1+\beta_1}\bigr)\cdot\bigl[1-\sum_{n=1}^N\frac{\tau_n}{\lambda_n(1-\frac{1+\beta_n}{\beta_n}\xi)}\bigr] & 0 & \ldots & 0 \\
                                \frac {\beta_2\tau_1}{\lambda_2(1+\beta_2)} & \xi-\frac
{\beta_2}{1+\beta_2} & \ldots & 0 \\
                                \vdots & \vdots & \ddots & \vdots \\
                                \frac {\beta_N\tau_1}{\lambda_N(1+\beta_N)} & 0 & \ldots & \xi-\frac
{\beta_N}{1+\beta_N} \\
                              \end{array}
                            \right|. \end{split}\end{displaymath}
Therefore, we can see that, the eigenvalues of $\textbf{J}^{BR}$ are
the roots of
\begin{equation}
\label{eq:eqn34}
\Bigl[\sum_{n=1}^N\frac{\tau_n}{\lambda_n(1-\frac{1+\beta_n}{\beta_n}\xi)}-1\Bigr]\cdot\prod_{n=1}^N\bigl(\xi-\frac{\beta_n}{1+\beta_n}\bigr)=0.
\end{equation}

Denote
$q(\xi)=\sum_{n=1}^N\frac{\tau_n}{\lambda_n(1-\frac{1+\beta_n}{\beta_n}\xi)}$.
First, we assume that $\beta_i\neq\beta_j, \forall i,j$. Without
loss of generality, consider $\beta_1<\beta_2<\cdots<\beta_N$. In
this case, the eigenvalues of $\textbf{J}^{BR}$ are the roots of
$q(\xi)=1$. Note that $q(\xi)$ is a continuous function and it
strictly increases in $(-\infty,\frac{\beta_1}{1+\beta_1})$,
$(\frac{\beta_1}{1+\beta_1},\frac{\beta_{2}}{1+\beta_{2}})$,
$\cdots$,
$(\frac{\beta_{N-1}}{1+\beta_{N-1}},\frac{\beta_{N}}{1+\beta_{N}})$,
and $(\frac{\beta_N}{1+\beta_N},+\infty)$. We also have
$\lim_{\xi\rightarrow(\frac{\beta_n}{1+\beta_n})^-}q(\xi)=+\infty$,
$\lim_{\xi\rightarrow(\frac{\beta_n}{1+\beta_n})^+}q(\xi)=-\infty,
n=1,2,\cdots,N$, and
$\lim_{\xi\rightarrow-\infty}q(\xi)=\lim_{\xi\rightarrow+\infty}q(\xi)=0$.
Therefore, the roots of $q(\xi)=1$ lie in
$(-\infty,\frac{\beta_1}{1+\beta_1})$,
$(\frac{\beta_1}{1+\beta_1},\frac{\beta_{2}}{1+\beta_{2}})$,
$\cdots$,
$(\frac{\beta_{N-1}}{1+\beta_{N-1}},\frac{\beta_{N}}{1+\beta_{N}})$.
Since $q(\xi)$ strictly increases in
$(-\infty,\frac{\beta_1}{1+\beta_1})$, we have $|\xi_n^{BR}|<1,
\forall n \in \mathcal{N}$ if and only if
$q(-1)=\sum_{n=1}^N\frac{\tau_n\beta_n}{\lambda_n(1+2\beta_n)}<1$.

Second, we consider the cases in which there exists
$\beta_i=\beta_j$ for certain $i,j$. Suppose that
$\{\beta_n\}_{n=1}^{N}$ take $K$ discrete values
$\kappa_1,\cdots,\kappa_K$ and the number of $\{\beta_n\}_{n=1}^{N}$
that equal to $\kappa_k$ is $n_k$. In this case, Equation
(\ref{eq:eqn34}) is reduced to
\begin{equation}
\label{eq:eqn35}
\Bigl[\sum_{n=1}^N\frac{\tau_n}{\lambda_n(1-\frac{1+\beta_n}{\beta_n}\xi)}-1\Bigr]\cdot\prod_{k=1}^K\bigl(\xi-\frac{\kappa_k}{1+\kappa_k}\bigr)^{n_k}=0.
\end{equation}
Hence, equation $q(\xi)=1$ has $N+K-\sum_{k=1}^Kn_k$ roots in total,
and $\xi=\frac{\kappa_k}{1+\kappa_k}$ is a root of multiplicity
$n_k-1$ for Equation (\ref{eq:eqn35}), $\forall k$. All these roots
are the eigenvalues of matrix $\textbf{J}^{BR}$. Similarly, the
roots of $q(\xi)=1$ lie in $(-\infty,\frac{\kappa_1}{1+\kappa_1})$,
$(\frac{\kappa_1}{1+\kappa_1},\frac{\kappa_{2}}{1+\kappa_{2}})$,
$\cdots$,
$(\frac{\kappa_{K-1}}{1+\kappa_{K-1}},\frac{\kappa_{K}}{1+\kappa_{K}})$.
A necessary and sufficient condition under which $|\xi_n^{BR}|<1,
\forall n \in \mathcal{N}$ is still $q(-1)<1$, i.e.
$\sum_{n=1}^N\frac{\tau_n\beta_n}{\lambda_n(1+2\beta_n)}<1$.
$\blacksquare$

\begin{remark} \label{rm:rm2}
Theorem \ref{th:th4} indicates that, if the condition in
(\ref{eq:eqn33}) is satisfied, the best response dynamics converges
linearly to the CE. The convergence rate is mainly determined by
$\max_{n \in \mathcal{N}}|\xi_n^{BR}|$. Suppose
$\beta_1<\beta_2<\cdots<\beta_N$ and
$\xi_1^{BR}<\xi_2^{BR}<\cdots<\xi_N^{BR}$. From the proof of Theorem
\ref{th:th4}, we can see that, under condition (\ref{eq:eqn33}),
$-1<\xi_1^{BR}<\frac{\beta_1}{1+\beta_1}<\xi_2^{BR}<\cdots<\xi_N^{BR}$,
and
$\frac{\beta_{N-1}}{1+\beta_{N-1}}<\xi_N^{BR}<\frac{\beta_{N}}{1+\beta_{N}}$.
Therefore, the rate of convergence can be approximated by $\max
\{|\xi_1^{BR}|,|\xi_N^{BR}|\}$. Note that choosing larger
$\{\lambda_n\}_{n=1}^{N}$ increases $\xi_1^{BR}$. Hence, if
$-1<\xi_1^{BR}<-|\xi_N^{BR}|$, increasing $\{\lambda_n\}_{n=1}^{N}$,
i.e. having more self-constraint users, accelerate the convergence
rate of the best response mechanism. On the other hand, since
$\xi_N^{BR}>\frac{\beta_{N-1}}{1+\beta_{N-1}}$, the convergence rate
is lower bounded by $\frac{\beta_{N-1}}{1+\beta_{N-1}}$. Therefore,
if more than two users associate large weighting factors $\beta$
with their individual actions in the utility functions, we have
$\frac{\beta_{N-1}}{1+\beta_{N-1}}\rightarrow 1$ and the best
response dynamics converges slowly.

%Note that increasing $\{\lambda_n\}_{n=1}^{N}$ will increase
%$\xi_1^{BR}$, and decreasing $\{\lambda_n\}_{n=1}^{N-1}$ or
%increasing $\lambda_N$ will increase $\xi_N^{BR}$. Hence, if
%$-1<\xi_1^{BR}<-|\xi_N^{BR}|$, increasing $\{\lambda_n\}_{n=1}^{N}$,
%i.e. having more self-constraint users, improves the convergence
%rate. Otherwise, the convergence rate is governed by $|\xi_N^{BR}|$,
%and increasing $\{\lambda_n\}_{n=1}^{N-1}$ or decreasing $\lambda_N$
%will improve the convergence rate.
\end{remark}

\begin{remark} \label{rm:rm3}
Theorem \ref{th:th4} generalizes the necessary and sufficient
condition derived in \cite{NE_RC}, where users are assumed to be
symmetric, i.e. $\tau_n=1, \forall n$ and they adopt the Nash
strategy by choosing $\lambda_n=\tau_n, \forall n$. Due to lack of
symmetry, the derivation in \cite{NE_RC} is not readily applicable
to analyze the convergence of the best response dynamics. The proof
of Theorem \ref{th:th4} instead directly characterizes the
eigenvalues of the Jacobian matrix, and hence, provides a more
general convergence analysis of the dynamic algorithms that allow
users to update their actions based on their independent linear
conjectures.
\end{remark}

\begin{remark} \label{rm:rm4}
In Type II games, a locally stable CE is also globally convergent,
which is purely due to the property of its utility functions
specified in (\ref{eq:eqn16}). From (\ref{eq:eqn32}), we can see
that all the elements in $\textbf{J}^{BR}$ are independent of the
joint play $\textbf{a}^{t-1}$. This is in contrast with Type I games
considered in \cite{Yi_JSAC}, where local stability of a CE may not
imply its global convergence and the best response dynamics may only
converge if the operating point is close enough to the steady-state
equilibrium.
\end{remark}

\subsubsection{Jacobi Update} We consider another alternative
strategy update mechanism called Jacobi update \cite{Jacobi_play}.
In Jacobi update, every user adjusts its action gradually towards
the best response strategy. At stage $t$, user $n$ chooses its
action according to
\begin{equation}
\label{eq:eqn36} a_n^t=J_n(\textbf{a}^{t-1}):=a_n^{t-1}+
\epsilon\bigl[B_n(\textbf{a}^{t-1})-a_n^{t-1}\bigr],
\end{equation}
in which the stepsize $\epsilon>0$ and $B_n(\textbf{a}^{t-1})$ is
defined in (\ref{eq:eqn31}). The following theorem establishes the
convergence property of the Jacobi update dynamics.
\begin{theorem}
\label{th:th5} In Type II games, for given
$\{\tau_n,\beta_n,\lambda_n\}_{n=1}^N$, the Jacobi update dynamics
converges if the stepsize $\epsilon$ is sufficiently small.
\end{theorem}

\emph{Proof}: The Jacobian matrix $\textbf{J}^{JU}$ of the
self-mapping function (\ref{eq:eqn36}) satisfies
$\textbf{J}^{JU}=(1-\epsilon)I+\epsilon\textbf{J}^{BR}$. Therefore,
its eigenvalues $\{\xi_n^{JU}\}_{n=1}^{N}$ are given by
$\xi_n^{JU}=1-\epsilon+\epsilon\xi_n^{BR}$. From the proof of
Theorem \ref{th:th4}, we know that $\xi_n^{BR}<1, \forall n\in
\mathcal{N}$. Therefore, if $\epsilon<\frac{2}{1-\min_n\xi_n^{BR}}$,
we have $\xi_n^{JU}\in(-1,1), \forall n\in \mathcal{N}$ and the
Jacobi update dynamics converges. $\blacksquare$

\begin{remark} \label{rm:rm5}
Theorem \ref{th:th5} indicates that, for any
$\{\tau_n,\beta_n,\lambda_n\}_{n=1}^N>0$, the Jacobi update
mechanism globally converges to a CE as long as the stepsize is set
to be a small enough positive number. In other words, the small
stepsize in the Jacobi update can compensate for the instability of
the best response dynamics even though the necessary and sufficient
condition in (\ref{eq:eqn33}) is not satisfied.
\end{remark}

\subsection{Stability of the Pareto Boundary}
In order to understand how to properly choose the parameters
$\{\lambda_n\}_{n=1}^N$ such that it leads to efficient outcomes, we
need to explicitly describe the steady-state CE in terms of the
parameters $\{\lambda_n\}_{n=1}^N$ of the belief functions. Denote
the joint action profile at CE as $(a^*_1,\ldots,a^*_N)$. From
Equation (\ref{eq:eqn31}), we know that
\begin{equation}
\label{eq:eqn37} (\lambda_n+\beta_n\tau_n)a_n^*+\sum_{m \in
\mathcal{N}\setminus\{n\}}\beta_n\tau_ma_m^*= \beta_n\mu, \forall n
\in \mathcal{N}.
\end{equation}
The solutions of the above linear equations are
\begin{equation}
\label{eq:eqn38}
a_n^{CE}=\frac{\beta_n\mu}{\lambda_n(1+\sum_{m=1}^N\frac{\tau_m\beta_m}{\lambda_m})},
\forall n \in \mathcal{N}.
\end{equation}
Based on the closed-form expression of the CE, the following theorem
indicates the stability of the Pareto boundary in Type II games.

\begin{theorem}
\label{th:th6} For Type II games, all the operating points on the
Pareto boundary are globally convergent CE under the best response
dynamics.
\end{theorem}

\emph{Proof}: Comparing Equations (\ref{eq:eqn21}) and
(\ref{eq:eqn38}), we can see that,
$(a^{CE}_1,\ldots,a^{CE}_N)=(a^{PB}_1,\ldots,a^{PB}_N)$ if and only
if $\lambda_n=\tau_n/\omega_n$. Substitute it into the LHS of
(\ref{eq:eqn33}):
\begin{equation}
\label{eq:eqn39}
\sum_{n=1}^N\frac{\tau_n\beta_n}{\lambda_n(1+2\beta_n)}=\sum_{n=1}^N\frac{\omega_n\beta_n}{1+2\beta_n}<\frac{\sum_{n=1}^N\omega_n}{2}=\frac{1}{2}.
\end{equation}
Condition (\ref{eq:eqn33}) is satisfied for all the Pareto-optimal
operating points. In fact, we have $\min_n\xi_n^{BR}=0$, which is
because
$q(0)=\sum_{n=1}^N\frac{\tau_n}{\lambda_n}=\sum_{n=1}^N\omega_n=1$.
Therefore, under the best response dynamics, the Pareto boundary is
globally convergent. $\blacksquare$

In addition, we also note that Theorem \ref{th:th5} already
indicates the stability of the Pareto boundary under Jacobi update
as long as the parameters $\{\tau_n,\beta_n,\lambda_n\}_{n=1}^N$ are
properly chosen.

\begin{remark} \label{rm:rm6}
Since $\sum_{n=1}^N\omega_n=1$, we can see from the previous proof
that, the belief configurations $\{\lambda_n\}_{n=1}^N$ lead to
Pareto-optimal operating points if and only if
\begin{equation}
\label{eq:eqn40} \sum_{n=1}^N\frac{\tau_n}{\lambda_n}=1.
\end{equation}
Therefore, we can see that, to achieve Pareto-optimality in these
non-cooperative scenarios, users need to choose the belief
parameters $\{\lambda_n\}_{n=1}^N$ to be greater than or equal to
the parameters $\{\tau_n\}_{n=1}^N$ in the utility function
$\{u_n\}_{n=1}^N$ and the summation of $\frac{\tau_n}{\lambda_n}$
should be equal to $1$. Define user $n$'s conservativeness as
$\frac{\tau_n}{\lambda_n}$, which reflects the ratio between the
immediate performance degradation $-\tau_n\Delta a_n$ in the actual
utility function and the long-term effect $-\lambda_n\Delta a_n$ in
the conjectured utility function if user $n$ increases its action by
$\Delta a_n$. The condition in Equation (\ref{eq:eqn40}) indicates
that, to achieve efficient outcomes, the non-collaborative users
need to jointly maintain moderate conservativeness by considering
the multi-user coupling and appropriately choosing
$\{\lambda_n\}_{n=1}^N$. By ``moderate", we mean that users are
neither too aggressive, i.e. $\lambda_n\rightarrow\tau_n$ and
$\sum_{n=1}^N\frac{\tau_n}{\lambda_n}\rightarrow N$, nor too
conservative, i.e. $\lambda_n\rightarrow+\infty$ and
$\sum_{n=1}^N\frac{\tau_n}{\lambda_n}\rightarrow 0$. If more than
one user plays the Nash strategy and choose $\lambda_n=\tau_n$,
Equation (\ref{eq:eqn40}) does not hold and the resulting operating
point is not Pareto-optimal. Therefore, myopic selfish behavior is
detrimental.

%As a special example, suppose user $n$ chooses its belief parameter
%according to $\lambda_n=\tau_n$ and plays the Nash strategy. For the
%resulting CE to be Pareto-optimal, we can see that, user $j$ needs
%to set its parameter $\lambda_j=+\infty$, $\forall j\in
%\mathcal{N}\setminus\{n\}$. In other words, if one user selfishly
%sets $\lambda_n=\tau_n$, all the remaining users should totally stop
%transmitting to ensure Pareto efficiency.
%
Similarly as in (\ref{eq:eqn22}), we have
\begin{equation}
\sum_{n=1}^N\omega_n\log\frac{u_n(\mathbf{a}^{CE})}{u_n(\mathbf{a}^{PB})}=\sum_{n=1}^N\omega_n\beta_n\log\frac{\tau_n(1+\sum_{j=1}^N\omega_j\beta_j)}{\lambda_n\omega_n(1+\sum_{j=1}^N\frac{\tau_j\beta_j}{\lambda_j})}+\log\frac{1+\sum_{j=1}^N\omega_j\beta_j}{1+\sum_{j=1}^N\frac{\tau_j\beta_j}{\lambda_j}}.
\end{equation}
Using Jensen's inequality, we can conclude
$\sum_{n=1}^N\omega_n\log\frac{u_n(\mathbf{a}^{CE})}{u_n(\mathbf{a}^{PB})}\leq0$
and
$\sum_{n=1}^N\omega_n\log\frac{u_n(\mathbf{a}^{CE})}{u_n(\mathbf{a}^{PB})}=0$
if and only if $\omega_n=\frac{\tau_n}{\lambda_n}, \forall n$.
Therefore, if a CE is Pareto efficient, user $n$'s conservativeness
$\tau_n/\lambda_n$ corresponds to the weight assigned to user $n$ in
the weighted proportional fairness defined in (\ref{eq:eqnop}).
\end{remark}

\begin{figure}
\centering
\includegraphics[width=0.6\textwidth]{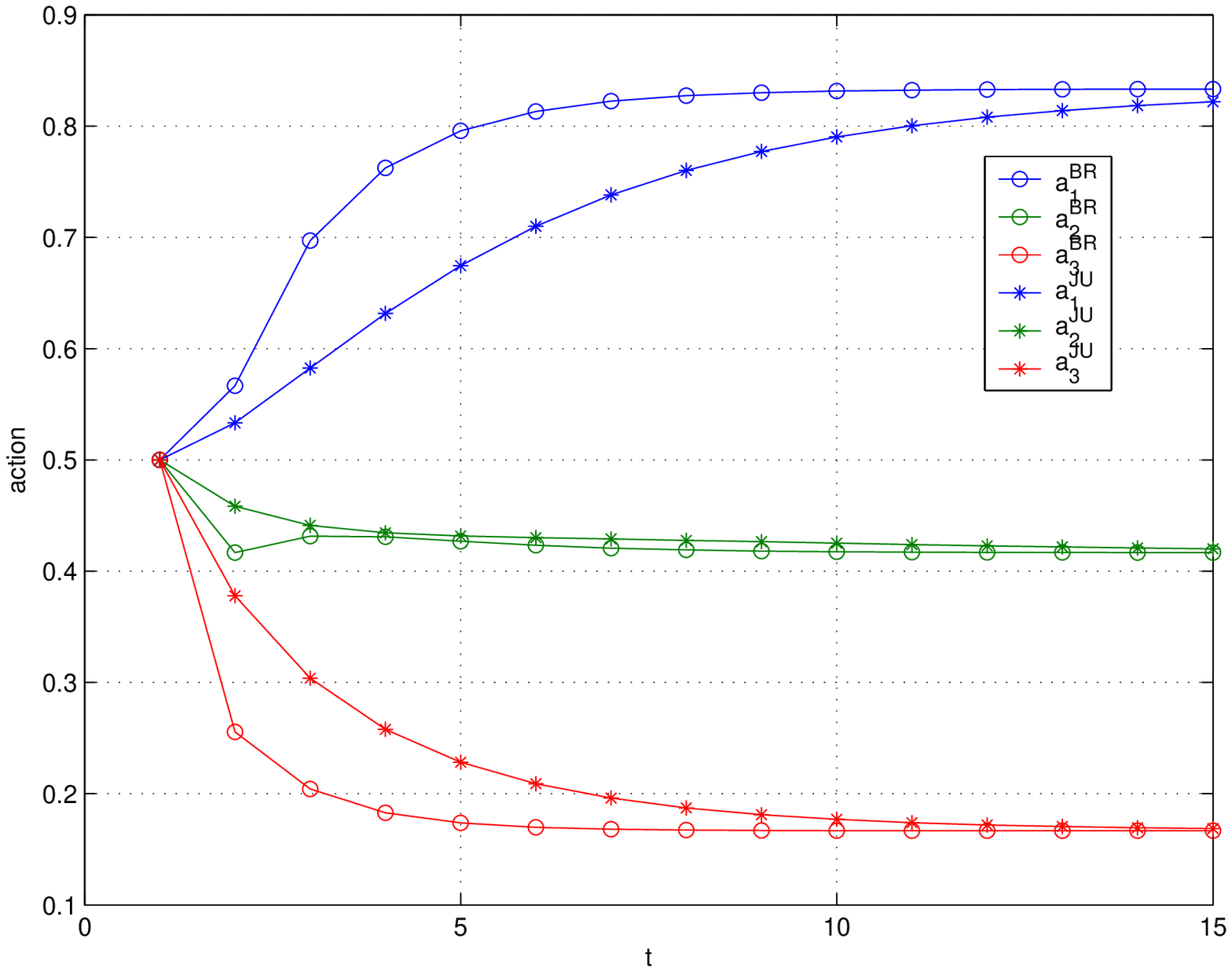}
\caption{The trajectory of the best response and Jacobi update
dynamics.} \label{fg:BR}
\end{figure}

As an illustrative example, we simulate a three-user system with
parameters $\mathbf{\beta}=[1.5 \; 1 \; 0.5], \; \mathbf{\tau}=[3 \;
4 \; 5], \mu=10, \omega_n=\frac{1}{3}, \forall n$. In this case, the
joint actions and the corresponding utilities at NE and Pareto
boundary are summarized in Table I. The price of anarchy quantified
according to (\ref{eq:eqn25}) is $-0.2877$ and the lower bound in
(\ref{eq:eqn25}) is $-0.5754$. As discussed in Section III.C, both
the upper bound and lower bound in (\ref{eq:eqn25}) are not tight.
Fig. \ref{fg:BR} shows the trajectory of the action updates under
both best response and Jacobi update dynamics, in which $a_n^0=0.5$,
$\lambda_n=\frac{\tau_n}{\omega_n}, \forall n$, and $\epsilon=0.5$.
The best response update converges to the Pareto-optimal operating
point in around 8 iterations and the Jacobi update experiences a
smoother trajectory and the same equilibrium is attained after more
iterations.

\subsection{Discussions}
\subsubsection{Comparison between Type I and Type II games}
As mentioned before, the properties of Type I games have been
investigated in the context of wireless random access\cite{Yi_JSAC}.
Table II summarizes some similarities and differences between both
types of games. First, the two algorithms exhibit different
properties under the best response dynamics. In Type I games, the
stable CE may not be globally convergent. However, the local
stability of a CE implies its global convergence in Type II games.
Second, it is shown in \cite{Yi_JSAC} that any operating point that
is arbitrarily close to the Pareto boundary of the utility region of
Type I games is a stable CE. Similarly, the entire Pareto boundary
of Type II games is also stable. At last, different relationships
between the parameter selection and the achieved utility at
equilibrium have been observed for the two types of games. In
particular, in Type I games, user $n$'s utility $u_n$ is
approximately proportional to the inverse of the parameter
$\lambda_n$ in its belief function. In contrast, in Type II games,
if the CE is Pareto-optimal, the ratio $\tau_n/\lambda_n$ coincide
with the weight $\omega_n$ assigned to user $n$ in the proportional
fairness objective function. In other words, based on the definition
of proportional fairness \cite{pf}, we know
\begin{equation}
\label{eq:eqn41} \sum_{n=1}^N\frac{\tau_n(u_n'-u_n^*)}{\lambda_n
u_n^*}\leq0,
\end{equation}
in which $(u_1',u_2',\ldots,u_N')$ is the users' achieved utility
associated with any other feasible joint action and
$(u_1^*,u_2^*,\ldots,u_N^*)$ is the optimal achieved utility for
problem (\ref{eq:eqnop}) with $\omega_n=\tau_n/\lambda_n$ and
$\sum_{n=1}^N\omega_n=1$.

\subsubsection{Pricing Mechanism vs. Conjectural Equilibrium}
In order to achieve Pareto-optimality, information exchanges among
users is generally required in order to collaboratively maximize the
system efficiency. The existing cooperative communication scenarios
either assume that the information about all the users is gathered
by a trusted moderator (e.g. access point, base station, selected
network leader etc.), to which it is given the authority to
centrally divide the available resources among the participating
users, or, in the distributed setting, users exchange price signals
(e.g. the Lagrange multipliers for the dual problem) that reflect
the ``cost" for consuming per unit constrained resources to maximize
the social welfare and reach Pareto-optimal allocations. As an
important tool, the pricing mechanism has been applied in the
distributed optimization of various communication networks
\cite{Num}. However, we would like to point out that, the pricing
mechanism generally requires repeated coordination information
exchange among users in order to determine the optimal actions and
achieve the Pareto-optimality. In contrast, for the linear coupled
communication games, since the specific structure of the utility
function is explored, the CE approach is able to calculate the
Pareto efficient operating point in a distributed manner, without
any real-time information exchange among users. In fact, the
underlying coordination is implicitly implemented when the
participating users initialize their belief parameters. Once the
belief parameters are properly initialized by the protocol according
to (\ref{eq:eqn40}), using the proposed dynamic update algorithms,
individual users are able to achieve the Pareto-optimal CE solely
based on their individual local observations on their states and no
message exchange is needed during the convergence process.
Therefore, the conjecture equilibrium approach is an important
alternative to the pricing-based approach in the linearly coupled
games.

\section{Conclusion}
We derive the structure of the utility functions in the multi-user
communication scenarios where a user's action has proportionally the
same impact over other users' utilities. The performance gap between
NE and Pareto boundary of the utility region is explicitly
characterized. To improve the performance in non-cooperative cases,
we investigate a CE approach which endows users with simple linear
beliefs which enables them to select an equilibrium outcome that is
efficient without the need of explicit message exchanges. The
properties of the CE under both the best response and Jacobi dynamic
update mechanisms are characterized. We show that the entire Pareto
boundary in linearly coupled games is globally convergent CE which
can be achieved by both studied dynamic algorithms without the need
of real-time message passing. A potential future direction is to see
how to extend the CE approach to the general linearly coupled games
that are compositions of the basic two types and certain particular
non-linearly coupled multi-user communication scenarios.

%\appendices
%\section{Proof of Theorem}

\ifCLASSOPTIONcaptionsoff
  \newpage
\fi

\newpage

\begin{table}
\centering \caption{Actions and payoffs at NE and Pareto boundary.}
\begin{tabular}{|c||c|c|c|} \hline
 & User 1 & User 2 & User 3 \\
\hline \hline $a_i^{NE}$ & $1.25$ & $0.625$ & $0.25$ \\
\hline $u_i^{NE}$ & $3.4939$ & $1.5625$ & $1.25$ \\
\hline $a_i^{PB}$ & $0.833$ & $0.417$ & $0.167$ \\
\hline $u_i^{PB}$ & $3.8036$ & $2.0833$ & $2.0412$ \\
\hline \end{tabular} \end{table}

\begin{table}
\centering \caption{Comparison between Type I and Type II games.}
\begin{tabular}{|c||c|c|c|} \hline
Games & Best response dynamics & Stability vs. efficiency & Fairness vs. parameter selection \\
\hline \hline
Type I & local stability $\Leftarrow$ global convergence & stable at near-Pareto-optimal points & $u_n \varpropto \tau_n/\lambda_n$\\
\hline
Type II & local stability $\Leftrightarrow$ global convergence & stable at the Pareto boundary  & $\omega_n=\tau_n/\lambda_n$ at the Pareto boundary\\
\hline\end{tabular}
\end{table}

\end{document}